\documentclass[prd,twocolumn,nofootinbib,superscriptaddress]{revtex4-1}
\usepackage{graphicx}
\usepackage{amssymb}
\usepackage{amsmath}
\usepackage{mathtools}
\usepackage{epstopdf}
\usepackage{aas_macros}
\usepackage{color}

\newcommand{\cpm}{{\small CUBEP$^3$M}}
\newcommand{\hfit}{{\small HALOFIT}}

\newcommand{\class}{{\small CLASS}}

\newcommand{\mpch}{\mbox{Mpc}/\mbox{h}}
\newcommand{\hmpc}{\mbox{h}/\mbox{Mpc}}
\newcommand{\ev}{{\rm eV}}
\newcommand{\mev}{{\rm meV}}

\newcommand{\kms}{{\rm km}/{\rm s}}

\begin{document}
\title{Measurement of the Cold Dark Matter-Neutrino Dipole in the TianNu Simulation}
 
\author{Derek Inman} \email{inmand@cita.utoronto.ca}
\affiliation{Canadian Institute for Theoretical Astrophysics,
  University of Toronto, M5S 3H8, Ontario, Canada}
\affiliation{Department of Physics, University of Toronto, Toronto,
  Ontario M5S 1A7, Canada}

\author{Hao-Ran Yu} \affiliation{Canadian Institute for Theoretical
  Astrophysics, University of Toronto, M5S 3H8, Ontario, Canada}
\affiliation{Kavli Institute for Astronomy \& Astrophysics, Peking
  University, Beijing 100871, China} \affiliation{Department of
  Astronomy, Beijing Normal University, Beijing 100875, China}

\author{Hong-Ming Zhu} \affiliation{Key Laboratory for Computational
  Astrophysics, National Astronomical Observatories, Chinese Academy
  of Sciences, Beijing 100012, China}

\author{J.D. Emberson} \affiliation{ALCF Division, Argonne National
  Laboratory, Lemont, IL 60439, USA}

\author{Ue-Li Pen} \email{pen@cita.utoronto.ca} \affiliation{Canadian
  Institute for Theoretical Astrophysics, University of Toronto, M5S
  3H8, Ontario, Canada} \affiliation{Dunlap Institute for Astronomy
  and Astrophysics, University of Toronto, Toronto, ON M5S 3H4,
  Canada} \affiliation{Canadian Institute for Advanced Research,
  Program in Cosmology and Gravitation} \affiliation{Perimeter
  Institute for Theoretical Physics, Waterloo, ON, N2L 2Y5, Canada}

\author{Tong-Jie Zhang} \email{tjzhang@bnu.edu.cn}
\affiliation{Department of Astronomy, Beijing Normal University,
  Beijing 100875, China} \affiliation{Shandong Provincial Key
  Laboratory of Biophysics, School of Physics and Electric
  Information, Dezhou University, Dezhou 253023, China}
\affiliation{National Supercomputer Center in Guangzhou, Sun Yat-Sen
  University, Guangzhou, 510275, China}

\author{Shuo Yuan} \affiliation{Department of Astronomy, Peking
  University, Beijing 100871, China}

\author{Xuelei Chen} \affiliation{Key Laboratory for Computational
  Astrophysics, National Astronomical Observatories, Chinese Academy
  of Sciences, Beijing 100012, China}

\author{Zhi-Zhong Xing} \affiliation{School of Physical Sciences,
  University of Chinese Academy of Sciences, Beijing 100049, China}
\affiliation{Institute of High Energy Physics, Chinese Academy of
  Sciences, Beijing 100049, China}

\begin{abstract}
  Measurements of neutrino mass in cosmological observations rely on
  two point statistics that are hindered by significant degeneracies
  with the optical depth and galaxy bias.  The relative velocity
  effect between cold dark matter and neutrinos induces a large scale
  dipole into the matter density field and may be able to provide
  orthogonal constraints to standard techniques.  We numerically
  investigate this dipole in the TianNu Simulation, which contains
  cold dark matter and 50 \mev{} neutrinos.  We first compute the
  dipole using a new linear response technique where we treat the
  displacement caused by the relative velocity as a phase in Fourier
  space and then integrate the matter power spectrum over redshift.
  Then, we compute the dipole numerically in real space using the
  simulation density and velocity fields.  We find excellent agreement
  between the linear response and N-body methods.  Utilizing the
  dipole as an observational tool will require two tracers of the
  matter distribution that are differently biased with respect to the
  neutrino density.
\end{abstract}

\maketitle

\begin{section}{Introduction}
  \label{sec:introduction}
  Terrestrial oscillation experiments have convincingly demonstrated
  that neutrinos are massive, and the measured mass splittings provide
  a lower bound on the neutrino mass scale:
  $M_\nu = \sum m_\nu \gtrsim 0.06$ \ev{} \cite{bib:Capozzi2016}.  The
  best upper bounds currently come from cosmological observations.
  The typical signature of cosmological neutrinos is a characteristic
  mass dependent suppression in the total matter power spectrum on
  small scales.  This provides a conceptually simple way to infer the
  neutrino mass from two-point statistics: measure the amplitude of
  fluctuations on large scales using the cosmic microwave background
  (CMB) and compare it to the small scale amplitude inferred from
  large scale structure observations.  For instance, Planck is
  sensitive to both the primary CMB (which depends on the scalar
  amplitude, $A_S$) and weak lensing of the CMB (which comes from
  smaller scale mass distributions), which yield a constraint of
  $M_\nu < 0.23$ \ev{} when combined \cite{bib:Planck2015}.
  Alternatively, the small scale measurement can be taken from direct
  large scale structure observations.  For example, combining Planck
  and Lyman-Alpha measurements from BOSS yield $M_\nu < 0.12$ \ev{}
  \cite{bib:Palanque-Delabrouille2015}.

  A number of upcoming experiments are aiming to improve this
  measurement of $M_\nu$ with the goal of resolving all possible
  masses down to the minimal mass of $\sim0.06$ \ev.  The Dark Energy
  Spectroscopic Instrument (DESI) is forecasted to have a mass
  resolution of $\sim 0.02$ eV \cite{bib:Font-Ribera2014} as does the
  next generation CMB Stage IV (CMB S4) experiment when combined with
  DESI baryon acoustic oscillation (BAO) measurements
  \cite{bib:CMBS42016}.  However, significant challenges remain in
  utilizing the two point function due to a number of parameters that
  must be precisely controlled.  The well known degeneracy between
  $A_S$ and the optical depth $\tau$ is still a consistent
  obstruction, and will hinder neutrino mass sensitivity in CMB S4 if
  the current measurement from Planck is not improved upon.  On small
  scales, the limiting factor is disentangling the small neutrino
  effect from large and uncertain baryonic physics.  To reach $0.02$
  \ev{} sensitivity, DESI must map its well resolved galaxy power
  spectrum to the underlying matter power spectrum which requires a
  highly precise knowledge of the galaxy bias. If this cannot be done,
  the DESI constraints with only BAO and Lyman-Alpha fall to $0.098$
  \ev.

  There is therefore clear motivation to search for new techniques
  beyond the two point function that may depend less, or differently,
  on $\tau$, the galaxy bias, or baryonic astrophysics.  A recent
  example of such a technique is differential neutrino condensation,
  which exploits fluctuations of the neutrino to CDM density ratio
  \cite{bib:Yu2016}.  In a series of papers we have studied another
  probe of neutrino mass that utilizes the relative velocity between
  the neutrinos and CDM.  In \cite{bib:Zhu2014}, a large scale
  CDM-neutrino dipole was predicted due to the displacement of
  initially concurrent regions of CDM and neutrino density.  In
  \cite{bib:Zhu2016}, a second smaller scale effect was identified as
  arising due to dynamical friction on CDM halos moving through a more
  homogeneous neutrino background.  Finally, \cite{bib:Inman2015}
  implemented neutrinos in the N-body code \cpm{}
  \cite{bib:HarnoisDeraps2013} and measured the relative velocity
  under non-linear gravitational evolution.  This relative velocity
  was also shown to be highly correlated with the underlying matter
  density field and can therefore be accurately predicted.
  
  In this paper we utilize the TianNu simulation, described in \S
  \ref{sec:simulation}, to compute the CDM-neutrino dipole in two
  different ways.  In \S \ref{ssec:response}, we compute the dipole
  via an integral over the non-linear CDM power spectrum with the
  relative velocity contributing a phase shift in Fourier space.  In
  \S \ref{ssec:Numerical}, we perform the calculation in real space by
  displacing hierarchically averaged density and velocity fields. We
  find that the latter is significantly larger on all scales compared
  to the predictions in \cite{bib:Zhu2014}, but is well matched by the
  new response computation provided we take into account the relative
  velocity correlation scale.
\end{section}

\begin{section}{TianNu Simulation}
  \label{sec:simulation}
  The TianNu simulation is an N-body simulation containing both CDM
  and $50$ \mev{} neutrino particles performed on the Tianhe-2
  supercomputer \cite{bib:Yu2016,bib:Emberson2016}.  Simulations with
  neutrinos require compromise between two competing needs: sample
  variance and Poisson noise.  On small scales, the large thermal
  motion of neutrinos causes them to be severely dominated by Poisson
  noise.  This noise can be reduced through the use of a large number
  of particles and a smaller box size.  On the other hand, the largest
  modes in the box will be limited by sample variance so the box
  cannot be too small.  The TianNu simulation uses $6912^3$ CDM
  particles and $13824^3$ neutrino particles in a box of size $1200$
  \mpch.  This choice allows for scales $0.05 < k/($\hmpc$) < 0.5$ to
  be relatively unafflicted from either form of noise.

  In addition to the Poisson noise, the large thermal velocities also
  make simulating neutrinos difficult at high redshift.  For our
  simulation, we evolve the CDM alone from redshift $100$ to $5$ and
  then inject neutrinos.  The simulation is then evolved to redshift
  $0.01$.

  For the dipole computation, we are interested in the CDM and
  neutrino density and velocity fields on relatively large scales.  We
  compute $\delta_c$ and $\delta_\nu$ via the Nearest-Grid-Point
  method \cite{bib:Hockney1988}.  The relative velocity is the
  difference between the CDM and neutrino velocities:
  $\vec{v}_{c\nu} \equiv \vec{v}_c-\vec{v}_\nu$ computed by taking the
  average velocity of the particles in each cell.  We compute the
  fields with only $576^3$ cells ($\sim 2.1$ \mpch{} per cell per
  dimension).  This both reduces the necessary computational time as
  well as smoothes over small scale structure to improve the accuracy
  of the average velocity method.

  \begin{figure}[htbp]
    \begin{center}
      \includegraphics[width=0.5\textwidth]{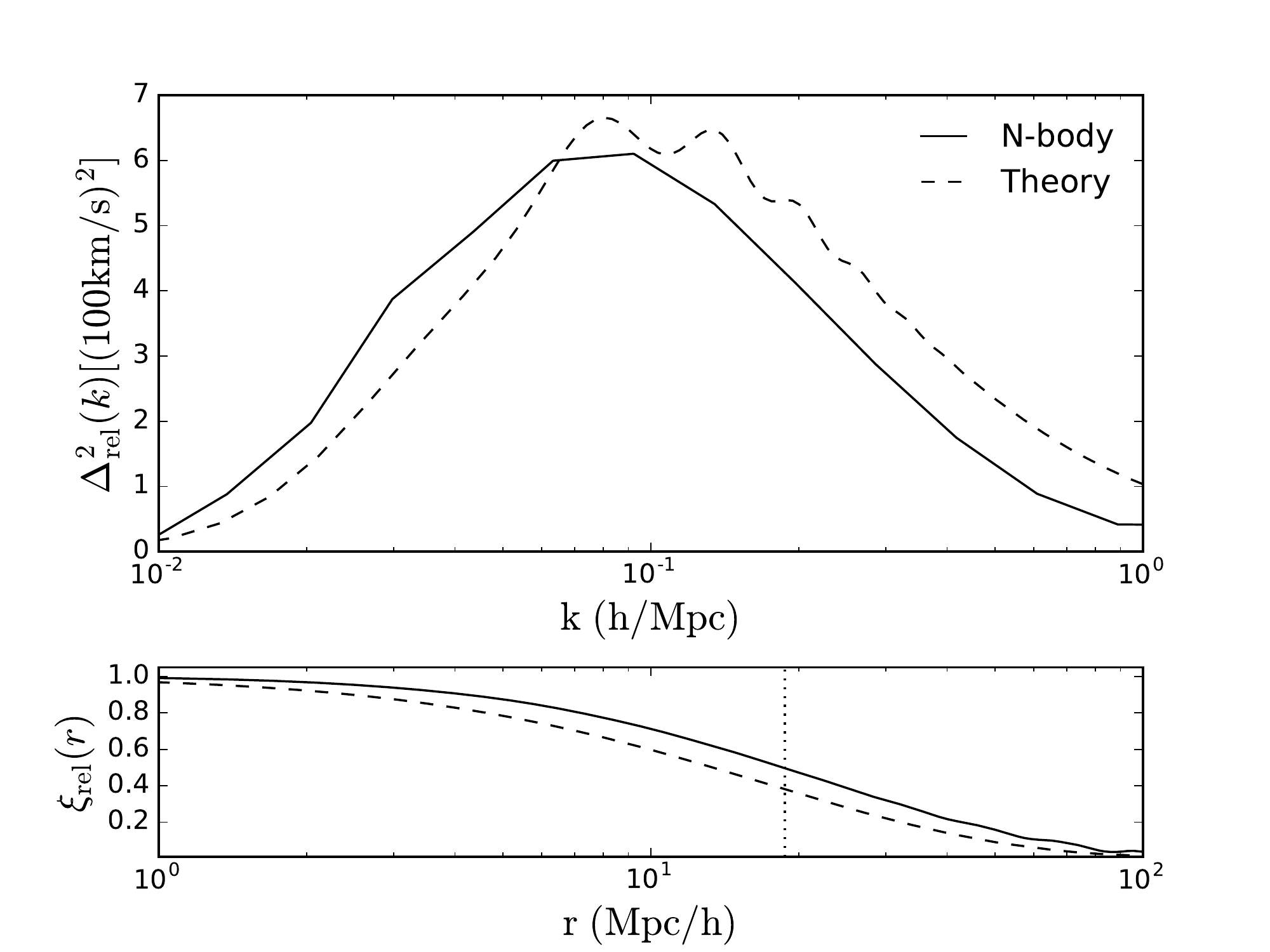}
      \caption{{\it Top.}  Cold dark matter-neutrino relative velocity
        power spectrum from the TianNu simulation (solid) and linear
        theory (dashed).  TianNu fields are computed on a reduced
        $576^3$ mesh using the average particle velocity in each
        cell. {\it Bottom.}  The corresponding correlation functions
        obtained by integrating the relative velocity power spectrum.
        The dotted vertical line is the simulation coherence length,
        the scale at which the correlation function drops to half its
        value.}
      \label{fig:relvelpow}
    \end{center}
  \end{figure}

  In the top panel of Fig. \ref{fig:relvelpow} we show the
  CDM-neutrino relative velocity power, $\Delta^2_{\text{rel}}(k)$.
  We find the N-body relative velocity to be slightly larger than that
  in \cite{bib:Inman2015} (the black curve in Fig. 10), especially on
  large scales.  This is likely due to the larger box size used
  ($1200$ \mpch{} versus $500$) which captures additional power from
  large scale modes.  We compute the integrated relative velocity:
  \begin{align}
    \label{eq:relvelnum}
    v_r=\sqrt{\int \Delta^2_{\rm rel}(k) \frac{dk}{k}} = 392\ \kms.
  \end{align}
  We also compute the monopole correlation function:
  \begin{align}
    \label{eq:moncorr}
    \xi(r) &= \int d^3k P(k) e^{(i\vec{k}\cdot\vec{r})} \nonumber \\
           &= \int dk k^2 P(k)/(2\pi^2) j_0(kr) \nonumber \\
    \xi(r) &\rightarrow \xi_0(r)/\xi_0(0)
  \end{align}
  with $P(k)$ being the monopole power spectrum,
  $\Delta^2(k) = \frac{k^3}{2\pi^2} P(k)$ and $j_0$ being the
  spherical Bessel function and the last line indicates that we divide
  by the $r=0$ value.  This result is shown in the bottom panel of
  Fig. \ref{fig:relvelpow}.  The correlation length is defined as the
  scale at which $\xi(R_{\rm rel})=0.5$.  We find $R_{\rm rel} = 18.7$
  \mpch{} in simulation, which is slightly larger than the linear
  value of $13.4$ \mpch.
\end{section}

\begin{section}{Dipole Correlation Function}
  \label{sec:dipcorrfun}
  \begin{subsection}{Linear Response Dipole}
    \label{ssec:response}
    In \cite{bib:Zhu2014}, the calculation of the dipole $P_{c\nu1}$
    was performed using Moving Background Perturbation Theory (MBPT).
    This approach, introduced in \cite{bib:Tseliakhovich2010}, solves
    the hydrodynamic Continuity and Euler Equations for CDM and
    neutrinos with the neutrinos having a constant sound speed
    proportional to their Fermi-Dirac dispersion.  Here, we use a new
    technique for describing the relative velocity that uses the
    linear response solution for the neutrino density field.  This
    allows for the full Fermi-Dirac distribution information to be
    utilized.

    Formally, neutrinos obey the non-linear collisionless Vlasov
    equation although fluid approximations are also in common use.
    Once linearized, the Vlasov equations has an integral solution
    (see e.g. \cite{bib:Ringwald2004, bib:AliHaimoud2013,
      bib:Inman2016} for additional discussions) of the form:
    \begin{align}
      \label{eq:integral}
      \delta_\nu = \frac{3}{2} H_0^2 \Omega_m \int_{s_i}^{s} ds' a
      (s-s') \delta_m(k,s') V(k(s-s')/\beta)
    \end{align}
    where $s$ is a time-like variable ($ds = a^2 dt$),
    $\delta_m \simeq \delta_c$ is the dominant density contrast and
    $\beta = m/(k_B T c)$ encodes the relevant neutrino properties:
    mass $m$ and temperature $T=(4/11)^{1/3}2.725$ K.  $V(x)$ is a
    function that encodes the neutrino's initial relativistic
    Fermi-Dirac velocity distribution:
    \begin{align}
      \label{eq:v0}
      V(x) = \frac{ \int dv v^2(\exp(v)+1)^{-1} j_0(xv) }{ \int dv v^2 (\exp(v)+1)^{-1}}.
    \end{align}
    The fluid approximation also has an integral solution with the
    same Eq. \ref{eq:integral} but $V(x)=j_0(x*c_s)$ where $c_s$ is
    the sound speed \cite{bib:Inman2016}.  We have verified that we
    are able to qualitatively reproduce the results in
    \cite{bib:Zhu2014} using the fluid linear response method.

    If neutrinos and CDM have a relative velocity $\vec{v}_r(z)$ that
    is coherent (i.e. independent of position $\vec{x}$), then the two
    species will flow past one another and become displaced
    $\vec{d} = \int \vec{v}_r(z) d\eta$ with $\eta$ the conformal
    time.  Since our simulation starts at such a late redshift, we
    simplify the displacement and take a constant relative velocity
    $\vec{v}_r(z) \simeq \vec{v}_r \rightarrow \vec{d}(\eta) =
    \vec{v}_r (\eta-\eta_i)$.
    In Fourier space, such a displacement leads to an additional
    phase, which we take to be in the CDM:
    $\delta_c(\vec{k}) \rightarrow
    \delta_c(\vec{k})e^{-i\vec{k}\cdot\vec{d}}$.
    Hence, the cross power can be written as
    $P_{c\nu} = \langle \delta_c \tilde{\delta}_\nu \rangle$ with
    \begin{align}
      \tilde{\delta}_\nu = &\frac{3}{2} H_0^2 \Omega_m \int_{s_i}^{s} ds' a
                             (s-s') \nonumber\\ &\delta_c(k,s') V(k(s-s')/\beta) e^{-i\vec{k}\cdot(\vec{d}(\eta)-\vec{d}(\eta'))}
    \end{align}
    with $\eta'$ corresponding to $s'$. If we define
    $\mu = \vec{k}\cdot\vec{d}/(kd)$ and expand the exponential factor
    $e^{-ikd\mu}\simeq 1 - ikd\mu$ we obtain the dipole component of
    the power spectrum,
    $(-i\mu)P_{c\nu1}(z,k,\mu) = -i \mu k v_r \langle \delta_c
    \tilde{\delta}_{\nu1} \rangle$:
    \begin{align}
      \label{eq:fourdipole}
      \tilde{\delta}_{\nu1} = &\frac{3}{2} H_0^2 \Omega_m \int_{s_i}^{s} ds' a
                                (s-s') \nonumber\\ &\delta_c(k,s') V(k(s-s')/\beta) (\eta-\eta')
    \end{align}
    where we have factored out the generic contributions of $\mu$,
    $v_r$ and $k$ from the integral.  For comparison with our
    numerical computation, we look for the antisymmetric combination:
    $P_{c\nu1}(\mu)-P_{c\nu1}(-\mu)$ which is twice that of
    Eq. \ref{eq:fourdipole}.

    \begin{figure}[htbp]
      \begin{center}
        \includegraphics[width=0.5\textwidth]{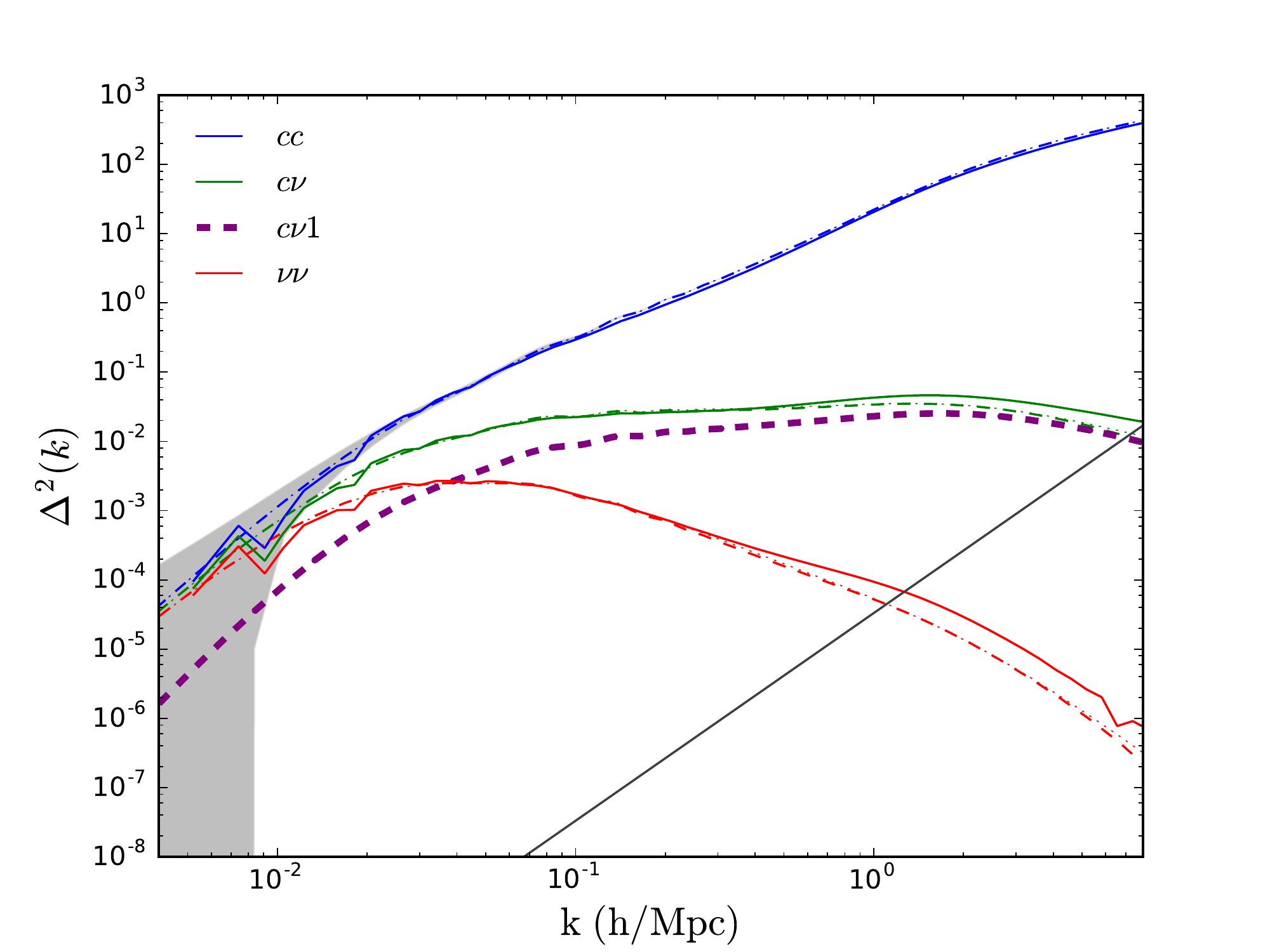}
        \caption{Cold dark matter (blue) and neutrino (red) auto and
          cross (green) power spectra in the TianNu simulation.  Solid
          curves are computed directly from the density fields.
          Dotted curves are computed via the \class{} Boltzmann code.
          Dashed curves utilize linear response.  The dipole cross
          power spectrum is shown in purple.  The straight line
          $\propto k^3$ and shown in dark grey is the neutrino Poisson
          noise.  The shaded grey region is the sample variance due to
          the simulation box size.}
        \label{fig:power}
      \end{center}
    \end{figure}

    We plot power spectra in Fig. \ref{fig:power}.  Solid lines are
    those computed from the TianNu simulation: blue is CDM, red is
    neutrino and green is the cross power.  Dashed curves are those
    from the linear response calculations described before, taking
    $\delta_c$ to be the square root of the HaloFit power spectrum
    $P_{HF}$ \cite{bib:Smith2003}.  Dotted curves are computed from
    the \class{} code \cite{ bib:Blas2011} assuming
    $P_{ij} = P_{HF}\frac{T_iT_j}{T_m^2}$ with $T_i$ being the linear
    transfer functions.  The filled grey region surrounding the
    \hfit{} power shows the sample variance of the simulation.  The
    dark grey curve crossing the neutrino spectrum is the Poisson
    noise (which we have subtracted out from the simulation neutrino
    power).  We see that there is a relatively noise free region as
    discussed earlier in \S \ref{sec:simulation}.

    The dipole correlation function is given by the Fourier transform
    of the dipole power spectrum:
    \begin{align}
      \label{eq:dipolepower}
      \xi_1(r) &= \int d^3k (-i\mu) P_1(k) e^{i\vec{k}\cdot\vec{r}} \nonumber \\
               &= \int dk k^2 P_1(k)/(2\pi^2) j_1(kr)
    \end{align}
    where $j_1(kr)$ is another spherical Bessel function.  In order to
    prevent ringing on large scales, we include a Gaussian cutoff,
    $\exp[-(k/(0.75\hmpc))^2]$, in the integral; since this cutoff
    suppresses power on small scales, we apply a high pass filter in
    $r$ to smoothly turn the cutoff on around $r\sim7.5$\mpch.  We
    perform this integral numerically and show the results in
    Fig. \ref{fig:response}.  We find significant enhancement on small
    scales compared to \cite{bib:Zhu2014} indicating that there is
    additional clustering that should be taken into account.
    \begin{figure}[htbp]
      \begin{center}
        \includegraphics[width=0.5\textwidth]{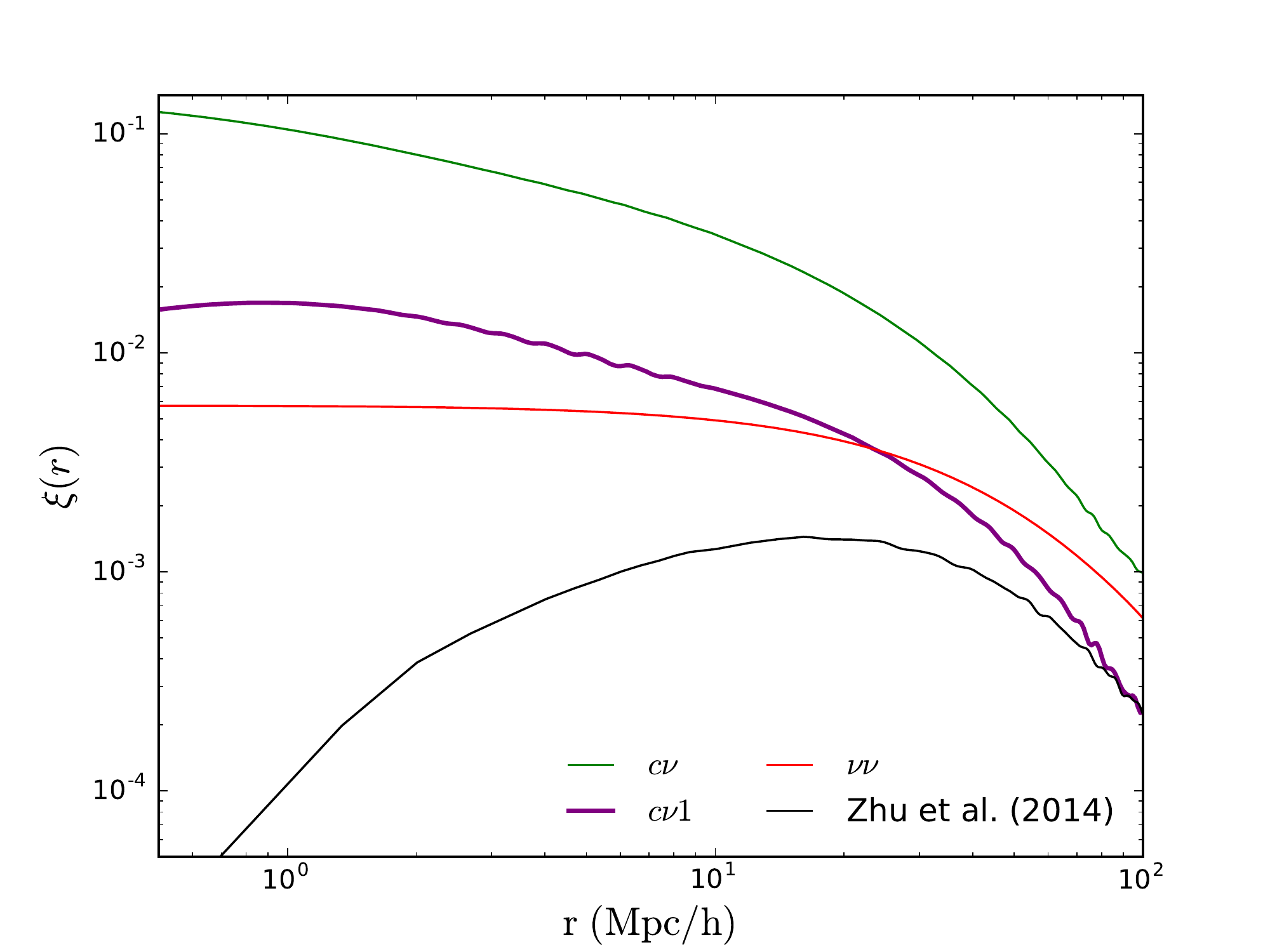}
        \caption{Correlation functions computed using the linear
          response method.  The green curve is the cold dark
          matter-neutrino cross correlation.  Red is the neutrino auto
          correlation function.  Purple is the dipole correlation
          function.  The black curve is reproduced from
          \cite{bib:Zhu2014} which approximated neutrinos as a perfect
          fluid with constant sound speed.}
        \label{fig:response}
      \end{center}
    \end{figure}
  \end{subsection}

  \begin{subsection}{Numerical Dipole}
    \label{ssec:Numerical}

    There are many ways to compute correlation functions
    \cite{bib:Zhang2005}.  We compute correlation functions in real
    space, using the method of hierarchical grids.  We first displace
    the grids and then multiply their elements.  This can be done in
    each direction (6 total displacements coresponding to
    $\pm x,\pm y,$ or $\pm z$), in each pair of directions (12 total
    displacements, e.g. $+x+y,-x+y,-x-y,+x-y$ and permutations with
    $z$) and for all three directions (8 total displacements, e.g. the
    four pairs of displacements discussed before with $\pm z$
    included).  After this, we average our grids down by a factor of
    two and repeat.  This algorithm is straightforward to parallelize
    which we have done with MPI, although it is not necessary for the
    computations here.  We note that averaging the grid to obtain
    larger displacements can cause a small amount of artifacting
    (after every third point).  This could be resolved by displacing
    the fields further rather than averaging, but this requires a
    significantly larger computational cost.

    We first compute the monopole correlation functions to validate
    our algorithm.  These results are shown in Fig. \ref{fig:monopole}
    alongside the linear response predictions.  We find excellent
    agreement at all scales as is expected.

    \begin{figure}[htbp]
      \begin{center}
        \includegraphics[width=0.5\textwidth]{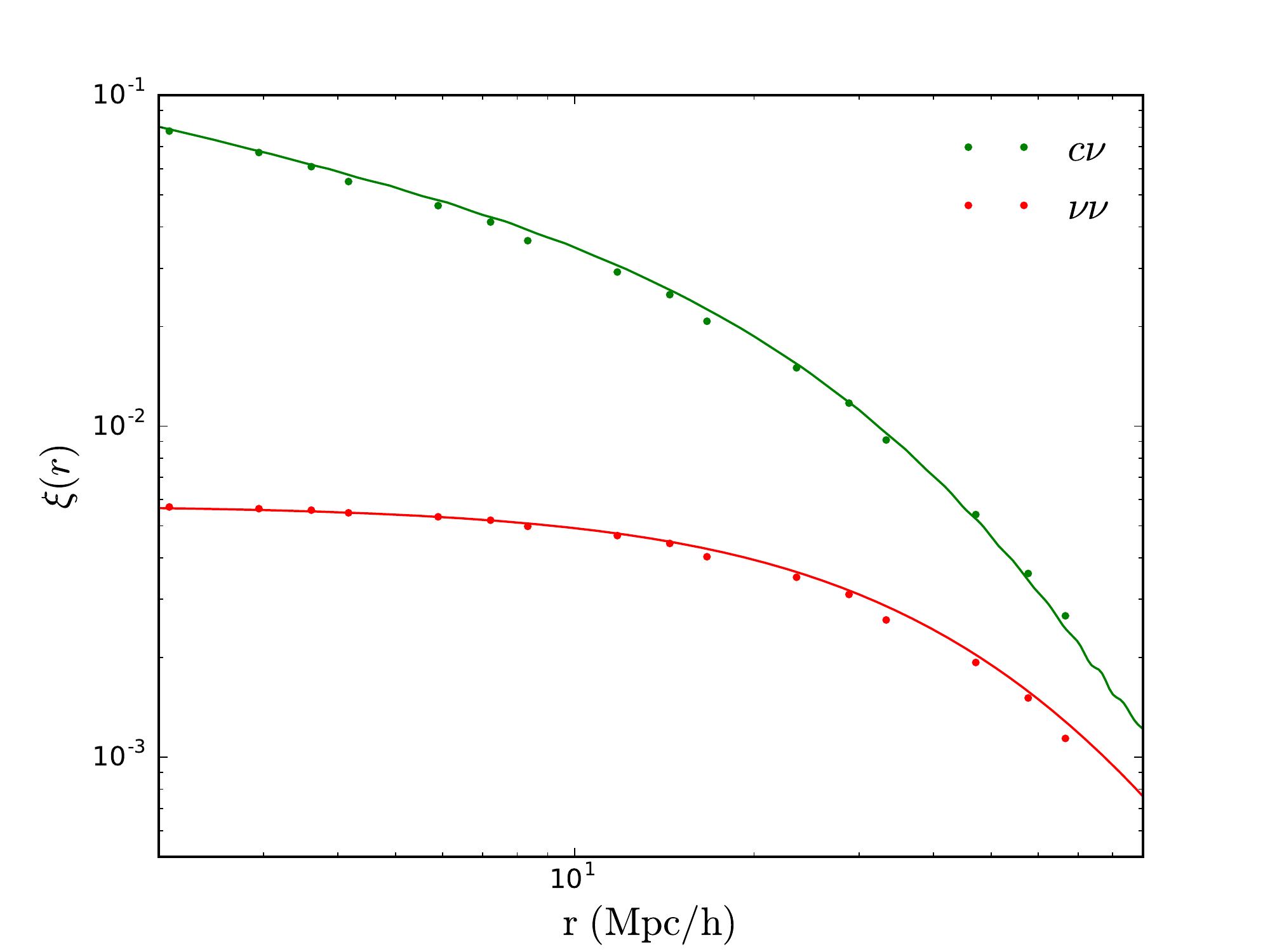}
        \caption{The monopole neutrino auto correlation function (red)
          and CDM-neutrino cross correlation function (green).  Dots
          are computed numerically in real space.  Lines are
          reproduced from Fig. \ref{fig:response} for comparison.}
        \label{fig:monopole}
      \end{center}
    \end{figure}

    We now describe the computation of the dipole directly from the
    TianNu simulation.  The relative velocity becomes a function of
    position, $\vec{v}_{\rm rel}(\vec{x})$, and we must now consider a
    three point correlation function:
    \begin{align}
      \label{eq:numdip1}
      \xi_{c\nu1} = \langle \delta_c(\vec{x})\delta_\nu(\vec{x}+\vec{r})
      \hat{v}_{\rm avg}(\vec{x},\vec{x}+\vec{r})\cdot\hat{r}\rangle
    \end{align}
    where we have the freedom to choose the functional dependence of
    $\hat{v}_{\rm avg}$ on $\vec{x}$ and $\vec{r}$ as is convenient.
    Since we expect the relative velocity to be coherent on scales
    $r<R_{\rm rel}$, we simply take the average value of the relative
    velocity at both $\vec{x}$ and $\vec{x}+\vec{r}$.  To see that
    this yields a dipole, consider the transformation
    $\vec{r}\rightarrow-\vec{r}$ and the change of variables
    $\vec{x}\rightarrow\vec{x}-\vec{r}$.  Since
    $\hat{v}_{\rm avg} (\vec{x},\vec{x}+\vec{r}) = \hat{v}_{\rm avg}
    (\vec{x}+\vec{r},\vec{x})$,
    the correlation function then becomes
    $\langle \delta_c(\vec{x}+\vec{r})\delta_\nu(\vec{x}) \hat{v}_{\rm
      avg} (\vec{x}+\vec{r},\vec{x})\cdot(-\hat{r})\rangle$.
    If we average this result with Eq. \ref{eq:numdip1}, we obtain the
    antisymmetric CDM-neutrino dipole correlation function,
    $\xi_{c\nu1}$, in the direction of the relative velocity, as:
    \begin{align}
      \label{eq:numericaldipole}
      \xi_{c\nu1}(r) = \frac{1}{2} \langle
      \left[\delta_c(\vec{x})\delta_\nu(\vec{x}+\vec{r}) -
      \delta_\nu(\vec{x})\delta_c(\vec{x}+\vec{r}) \right]
      \hat{v}_{\rm avg} \cdot\hat{r} \rangle
    \end{align}
    with
    \begin{align}
      \vec{v}_{\rm avg} = \frac{1}{2}  \left(
      \vec{v}_{c\nu}(\vec{x}) + \vec{v}_{c\nu}(\vec{x}+\vec{r}) \right) 
    \end{align}
    where it is understood we should only displace our fields in one
    direction.  We see that generic CDM effects must cancel due to the
    antisymmetric combination of density fields.  In other words,
    ``$\xi_{cc1}$"$=0$.  Furthermore, correlations orthogonal to the
    relative velocity do not contribute.

    \begin{figure}[htbp]
      \begin{center}
        \includegraphics[width=0.5\textwidth]{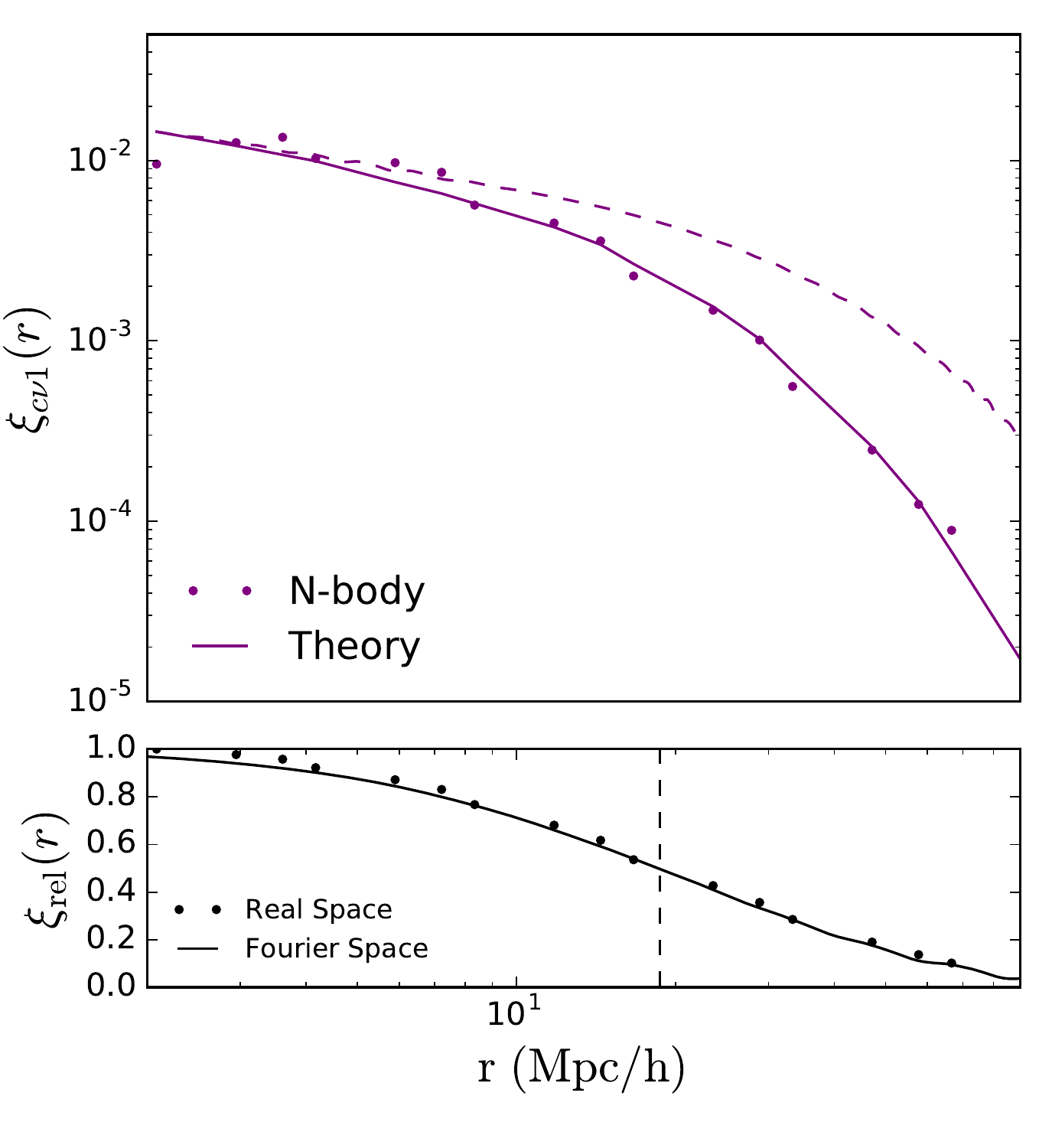}
      \caption{{\it Top.}  Cold dark matter-neutrino dipole
        correlation functions.  The dots are computed using
        hierarchical grid displacements of density and velocity
        fields.  The dashed curve is computed using linear response.
        The solid curve is the linear response multiplied by the
        velocity correlation function (black curve in lower panel).
        {\it Bottom.}  Relative velocity correlation function.  The
        dots are computed using hierarchical grid displacements.  The
        solid curve is computed as the Fourier transform of the
        relative velocity power spectrum.  Both are normalized by
        their $r=0$ value.  The dashed vertical line indicates the
        half-max velocity correlation length of $18.7$ \mpch.}
      \label{fig:dipole}
    \end{center}
  \end{figure}

  We show the CDM-neutrino dipole in the direction of the relative
  velocity as the purple points in the upper panel of
  Fig. \ref{fig:dipole}.  In addition we show the response dipole of
  Fig. \ref{fig:response} in dashed purple.  On the largest scales
  there is some disagreement. This is not a surprise: the MBPT
  approximation assumes a coherent flow over the entire box.  In
  reality, it is only coherent to scales $r\lesssim R_{\rm rel}$ and
  we therefore expect a smaller signal at distances larger than the
  coherence length.  Fortunately, the information on the coherency of
  the relative velocity is precisely the relative velocity correlation
  function, $\xi_{\rm rel}(r)$, which we show in the lower panel of
  Fig. \ref{fig:dipole}.  Thus, by multiplying the MBPT prediction by
  $\xi(r)$ we obtain the solid purple curve in the upper panel of
  Fig. \ref{fig:dipole} which matches simulations remarkably well.
\end{subsection}
\end{section}

\begin{section}{Discussion and Conclusion}
  \label{sec:discussion}
  We have demonstrated that the CDM-neutrino dipole exists and is well
  predicted via linear response for $50$ \mev{} mass neutrinos.
  However, the real space computation performed here relies on knowing
  both the neutrino density and relative velocity.  In
  \cite{bib:Inman2015} it was demonstrated that the relative velocity
  direction is predictable with either a CDM or halo density field.
  In fact, the neutrino density field is also predictable
  \cite{bib:Yu2016}.  However, this is {\it assuming} you know the
  transfer function.  For a massive neutrino universe this transfer
  function can be computed from a Boltzmann code; however, for a
  massless neutrino Universe it is zero.  Hence, while we can predict
  both the relative velocity and neutrino density at the same time,
  this does not result in an observable quantity.

  In \cite{bib:Zhu2014}, the proposed observable was to use two
  distinct tracers of the density field that are differentially
  biased with respect to neutrinos.  For instance, halos (observed via
  proxy galaxies) above and below a certain mass.  A second option
  would be the lensing field (which depends on the linear sum of CDM
  and neutrino densities) and a galaxy survey (which depends primarily
  on CDM).  We intend to investigate this in subsequent works.
\end{section}

\acknowledgements{
  \label{sec:acknowledgements}
  The TianNu and TianZero simulations were performed on Tianhe-2
  supercomputer at the National Super Computing Centre in Guangzhou,
  Sun Yat-Sen University.  This work was supported by the National
  Science Foundation of China (Grants No. 11573006, 11528306,
  10473002, 11135009), the Ministry of Science and Technology National
  Basic Science program (project 973) under grant No. 2012CB821804,
  the Fundamental Research Funds for the Central Universities.  Parts
  of the analysis were performed on the General Purpose Cluster
  supercomputer at the SciNet HPC Consortium \cite{bib:Loken2010}.
  SciNet is funded by: the Canada Foundation for Innovation under the
  auspices of Compute Canada; the Government of Ontario; Ontario
  Research Fund - Research Excellence; and the University of Toronto.
  We thank Prof. Yifang Wang of IHEP for his great initial support for
  our project, and Prof. Xue-Feng Yuan for his kindly great support in
  Tianhe-2 supercomputing center. We thank Joel Meyers for valuable
  discussions.  Authors DI, HRY, and ULP acknowledge the support of the
  NSERC.  HRY acknowledges General Financial Grant No.2015M570884 and
  Special Financial Grant No.2016T90009 from the China Postdoctoal
  Science Foundation.  XC acknowledges MoST 863 program 2012AA121701,
  CAS grant QYZDJ-SSW-SLH017, and NSFC grant 11373030. }

\bibliographystyle{apsrev}
\bibliography{thebib}

\end{document}